\documentclass[twocolumn,prl,longbibliography,superscriptaddress]{revtex4-1}
\usepackage{amsmath}
\usepackage{graphicx}
\usepackage{soul,xcolor}
\setstcolor{red}
\definecolor{mycolor}{rgb}{0.1, 0.1, 0.7}
\usepackage{dsfont}
\usepackage{url}
\usepackage{soul}
\usepackage[colorlinks]{hyperref}

\hypersetup{
 plainpages=true,
 breaklinks=true,
 hypertexnames=false,
 pageanchor=true,
 colorlinks=true,
 linkcolor={mycolor},
 citecolor={mycolor},
 urlcolor={mycolor},
 anchorcolor={black}
 }
\newcommand{\ket}[1]{|{#1}\rangle}

\def\doubleunderline#1{\underline{\underline{#1}}}

\begin{document}
\title{Self-stimulated pulse echo trains from inhomogeneously broadened spin ensembles 
}
\author{Kamanasish Debnath}
\email[e-mail:]{kamanasish.debnath@phys.au.dk}
\affiliation{Department of Physics and Astronomy, Aarhus University,
Ny Munkegade 120, DK-8000, Aarhus C, Denmark}
\author{Gavin Dold}
\affiliation{London Centre for Nanotechnology, University College London, London WC1H 0AH, United Kingdom}
\affiliation{National Physical Laboratory, Hampton Road, Teddington TW11 0LW, United Kingdom}
\author{John J. L. Morton}
\affiliation{London Centre for Nanotechnology, University College London, London WC1H 0AH, United Kingdom}
\affiliation{Department of Electronic and Electrical Engineering, UCL, London WC1E 7JE, United Kingdom}
\author{Klaus M{\o}lmer} 
\affiliation{Department of Physics and Astronomy, Aarhus University,
Ny Munkegade 120, DK-8000, Aarhus C, Denmark}
\date{\today}

\begin{abstract}
We show experimentally and describe theoretically how a conventional magnetic resonance Hahn echo sequence can lead to a self-stimulated pulse echo train when an inhomogeneously broadened spin ensemble is coupled to a resonator. Effective strong coupling between the subsystems assures that the first Hahn echo can act as a refocussing pulse on the spins, leading to self-stimulated secondary echoes. Within the framework of mean field theory, we show that this process can continue multiple times leading to a train of echoes. We introduce an analytical model that explains the shape of the first echo and numerical results that account well for the experimentally observed shape and strength of the echo train and provides insights into the collective effects involved.
\end{abstract}
\maketitle
\emph{Introduction--} Electron spin resonance (ESR)~\cite{PhysRev.115.1546, PhysRevLett.118.037701} and nuclear magnetic resonance (NMR)~\cite{Aslam67, RevModPhys.68.855} are used in diverse branches of science, ranging from spectroscopic studies in
biochemistry and materials science~\cite{doi:10.1021/,doi:10.1021/ar500340a,FAN201618,doi:10.1002/aenm.201602226} to imaging of internal organs in medicine~\cite{mni_nature}. 
In NMR and ESR, an ensemble of spins is typically placed within a resonator, controlled by the application of resonant pulses, and measured via emission of signals into a resonator mode. As the spin ensembles are typically inhomogeneous, a common solution is to use control pulses which refocus inhomogeneous interactions, reversing the time evolution of different spin packets to produce a spin echo~\cite{PhysRevLett.13.567, PhysRevLett.30.158} or `Hahn echo'~\cite{PhysRev.80.580}. 
Aside from being the cornerstone of pulsed NMR and ESR techniques, spin echoes have also become an essential ingredient in quantum information science  due to their applications in ensemble quantum memories with ESR and optical transitions~\cite{PhysRevLett.105.140503,PhysRevA.88.062324, PhysRevLett.110.250503,PhysRevX.4.021049, Simon2010}, and in nanoscale quantum metrology~\cite{nature_sensisng,PhysRevX.7.041011}, as well as being a building block for more complex dynamical decoupling sequences used to extend qubit coherence times~\cite{bargil2012}.

\begin{figure}
\centering
\includegraphics[width=0.45\textwidth]{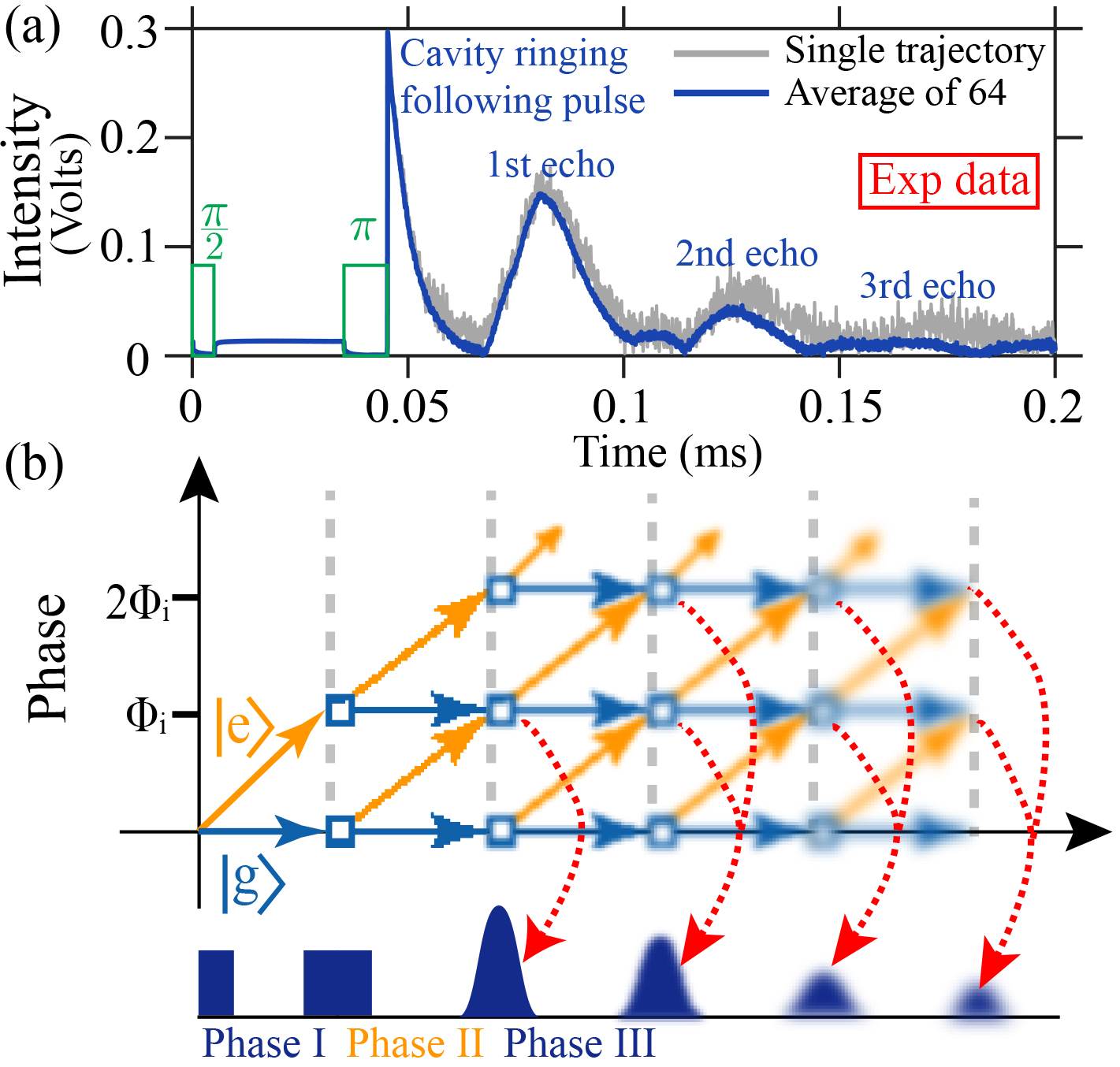}
\caption{(a) Experimentally observed emission from an inhomogeneously broadened spin ensemble ($^{145}$Nd ions doped at 200~ppm in a Y$_2$SiO$_5$ host) in a cavity subject to a conventional Hahn echo  sequence with $\tau= 30$ $\mu$s pulse delay. The spin-cavity coupling exhibits high cooperativity $C=153$ (see Supplementary Material~\cite{Supp} for full details). For reference, the $\pi/2$ and $\pi$ pulses are shown in green on a different intensity scale. (b) Schematic of the refocussing mechanism leading to a self-stimulated spin echo train (see text).}
\label{F1}
\end{figure}

The Hahn echo sequence consists of an initial $\pi/2$ pulse, a time interval $\tau$ and a $\pi$ pulse leading to the emission of a spin echo at time $2\tau$. The $\pi/2$ pulse excites the spins into a coherent superposition state, which starts precessing at the Larmor frequencies of the individual spins, leading to dephasing of the collective spin. At time $\tau$, a $\pi$ rotation of the spins is performed in the Bloch sphere, which is equivalent to a time reversal operation, since the phases of the excited and ground state amplitudes get interchanged, i.e. $\mathcal{A}\ket{{\rm g}} + \mathcal{B} e^{-i\Phi}\ket{{\rm e}} \rightarrow \mathcal{B} e^{-i\Phi}\ket{{\rm g}} + \mathcal{A}\ket{{\rm e}}$, where $\Phi$ is the relative phase between the ground state $\ket{{\rm g}}$ and the excited state $\ket{{\rm e}}$. As $\Phi\propto\omega^j_a\tau$, where $\hbar\omega^j_a$ is the energy difference between $\ket{e}$ and $\ket{g}$, the subsequent time evolution leads to refocussing of the spins to recover the original coherent superposition state and produce a spin echo at time $2\tau$. 

A range of recent ESR studies have begun to examine increasing the coupling between the spin ensemble and the resonator, for example to improve spin sensitivity in spectroscopic applications~\cite{bienfait2015,probst2017}, or to improve the efficiency of a quantum memories or transducers~\cite{PhysRevApplied.11.054082,PhysRevLett.105.140503,PhysRevA.88.062324,PhysRevA.98.063815, PhysRevLett.110.250503,PhysRevX.4.021049, Simon2010, PhysRevLett.107.060502,PhysRevA.85.053806,PhysRev.93.99, PhysRevA.100.053821,PhysRevA.73.020302, PhysRevA.84.063810}. However, increasing the spin-resonator coupling also introduces the possibility for the emitted echo itself to act as a sufficient perturbing field to drive further evolution of the spins. If the light-matter coupling becomes stronger than the dissipative losses, even the simple and ubiquitous Hahn echo sequence can yield non-trivial spin dynamics which lead to the emission of not just to one, but multiple spin echoes. Such multiple echoes are illustrated in Fig.~\ref{F1}(a) which shows the experimentally observed emission from an inhomogeneously broadened spin ensemble (Nd ions in Y$_2$SiO$_5$), subject only to a Hahn echo sequence. (further details on the experiment are provided in the Supplementary Material~\cite{Supp}). The data reveal the conventional echo signal at time $2\tau$, followed by additional echoes separated by $\tau$. A similar result was reported recently in Ref.~\cite{arxivpreprint1}, which showed that the secondary echoes were absent when the light-matter coupling was reduced. Indeed, secondary echoes were discerned in the very first microwave spin echo experiment in 1958~\cite{PhysRevLett.1.368}, where highly doped samples were used to compensate low detector sensitivity.
In this Letter we examine the `echo train' phenomenon experimentally and theoretically, with the goal of developing a better understanding of the underlying mechanisms, necessary for the exploitation of strongly coupled resonators and spin ensembles.

Before we present our theoretical results, we offer an intuitive and qualitative description of the dynamics of the spins subjected to imperfect $\pi/2$ and $\pi$ pulses.  Assuming that a spin with frequency $\omega^j_a$ is initially in the ground state, application of the first pulse excites the spin to a superposition state  $\mathcal{A}\ket{{\rm g}} + \mathcal{B}\ket{{\rm e}}$. A free evolution for time $\tau$ leads to $\mathcal{A}\ket{{\rm g}} + \mathcal{B} e^{-i\omega^j_a\tau}\ket{{\rm e}}$. A schematic of the above process is shown in Phase I in Fig.~\ref{F1}(b), where blue(orange) corresponds to the phase acquired by the ground(excited) state. The second pulse updates the state of the spin to
\begin{align}
\mathcal{A}_{1}\ket{{\rm g}} + \mathcal{A}_2\ket{{\rm e}} + \mathcal{B}_1 e^{ -i\omega^j_a\tau}\ket{{\rm e}} + \mathcal{B}_2 e^{-i\omega^j_a\tau}\ket{{\rm g}}.
\label{E1}
\end{align}
Here and in the following, the state amplitudes $\mathcal{A},\mathcal{A}_{m},\mathcal{B},\mathcal{B}_{m}$ etc. explore different values that need not be specified for our qualitative discussion. In the schematic in Fig.~\ref{F1}(b), the action of each pulse is shown as bifurcating blue and orange arrows representing the phase evolution for time $\tau$ of the respective ground and excited state amplitudes. The figure also depicts how, after a subsequent time $\tau$, two terms in the time evolved state (underlined below) come in phase:
\begin{align}
\mathcal{A}_1\ket{{\rm g}} + \underline{\mathcal{A}_2 e^{-i\omega^j_a\tau}\ket{{\rm e}}} + \mathcal{B}_1 e^{-2i\omega^j_a\tau}\ket{{\rm e}} + \underline{\mathcal{B}_2 e^{-i\omega^j_a\tau}\ket{{\rm g}}}
\label{E2}
\end{align}
This rephasing occurs for all values of the frequency of the spins and leads to the usual Hahn echo. 

In case of a perfect $\pi$ refocussing pulse, the state amplitudes $\mathcal{A}_1$ and $\mathcal{B}_1$ vanish, and the spins accumulate diverging phases in the time following this spin echo. Otherwise, all four terms in \eqref{E2} contribute to the subsequent evolution of the system, and this is key for the production of further echoes. For large $N$, the Hahn echo pulse may be strong enough to significantly alter the spin states and thus populate a new superposition: 
\begin{align}
(\mathcal{A}_{11}\ket{{\rm g}} + \underline{\mathcal{A}_{12}\ket{{\rm e}})} + e^{ -i\omega^j_a\tau}(\underline{\mathcal{C}_{21}\ket{{\rm g}}} + \doubleunderline{\mathcal{C}_{22}\ket{{\rm e}}}) \nonumber \\
+ e^{-2i\omega^j_a\tau}(\doubleunderline{\mathcal{B}_{11}\ket{g}} + \mathcal{B}_{12}\ket{e}).
\label{E3}
\end{align}
As can be seen from Fig.~\ref{F1}(b), after a further time $\tau$, the underlined (and doubly underlined) terms come into phase and cause a second echo. The evolution and refocussing may 
occur also for terms with higher phase arguments, and the process can repeat and lead to multiple self-stimulated echoes. 

To assess the validity of our qualitative discussion, we now proceed to investigate whether the spin echo train can be reproduced by a numerical treatment. We begin with a description of the model, followed by the numerical results, demonstrating clear evidence of the self-stimulated spin echo train. We then introduce a simplified analytical model that predicts various shapes of the echo pulse and provide insights into the collective effects involved. 

\emph{Theoretical model--} We consider $N= 10^{10}$ $^{145}$Nd spins coupled to a resonator of linewidth $\kappa= 2\pi\times150$~kHz and resonance frequency $\omega_c$. The spins are inhomogeneously broadened with transition frequencies $\omega^j_a$, following a Gaussian distribution with central frequency $2\pi\times8$ GHz and FWHM of $\Gamma_{\rm inh}= 2\pi\times 4$~MHz.  An external coherent drive with frequency $\omega_p$ and amplitude $F$ is used to apply the two initial pulses. In the frame rotating with the pump frequency, the Hamiltonian ($\hbar=1$) of the system can be written as:
\begin{equation}
H= \delta_c a^{\dagger}a + \sum_{j=1}^N\Big[\frac{\delta^j_a}{2}\sigma^j_z + g_j(a\sigma^j_+ + a^{\dagger}\sigma^j_-)\Big] + F(a + a^{\dagger})
\label{E4}
\end{equation}
where $\delta_c= (\omega_c-\omega_p)$ and $\delta^j_a= (\omega^j_a - \omega_p)$. $\sigma^j_z, \sigma^j_+, \sigma^j_-$ are the Pauli operators and $a, a^{\dagger}$ are the annihilation and creation operators of the cavity mode, obeying the usual commutation relation $[a,a^{\dagger}]=1$. For numerical simplicity, we assume $g_j=g=2\pi\times 8$ Hz. Despite this weak single spin-cavity coupling, the presence of $10^{10}$ spins significantly enhances the effective coupling. The exact dynamics of the system can be described by the master equation $\dot{\rho}= -i[H, \rho] + \kappa\mathcal{D}[a]\rho + \gamma\sum_{j=1}^N\mathcal{D}[\sigma^j_-]\rho + \Gamma\sum_{j=1}^N\mathcal{D}[\sigma^j_z]\rho$. The superoperator is defined as: $\mathcal{D}[\mathcal{O}]\rho= \mathcal{O}\rho\mathcal{O}^{\dagger} - \frac{1}{2}\{\mathcal{O}^{\dagger}\mathcal{O}, \rho\}$, for the operators $\mathcal{O}= a, \sigma^j_-, \sigma^j_z$. The lifetime $\gamma^{-1}$ of the excited state of the spins is of the order of seconds, and it is a good approximation to neglect spin decay while spin dephasing with a rate $\Gamma \simeq 2\pi\times 1$~kHz plays a significant role on the time scale of interest. Solving the master equation for the full density matrix is impossible, and we treat the model by discretizing the frequency distribution into $N_k= 10^5$ frequency classes following a Gaussian distribution and employ mean field theory for the field and spin raising and lowering operators, assuming factorization of their products i.e. $\langle AB\rangle\approx \langle A\rangle\langle B\rangle$~\cite{PhysRevA.98.063837}.

\emph{Results--} We apply a strong classical square pulse of amplitude $F= 50$~GHz from $t_1= 0.20$~$\mu$s to $t_2= 0.42$~$\mu$s, followed by a delay of variable duration $\tau$ and a second pulse between $t_3= t_2 + \tau$ and $t_4= t_3 + 0.43$~$\mu$s. Any finite pulse area causes some coherent transfer of population between the spin eigenstates and the exact intensity is not crucial for the appearance of the echo as long as the pulses excite the spins  by a significant amount.

\begin{figure*}
\centering
\includegraphics[width=0.98\textwidth]{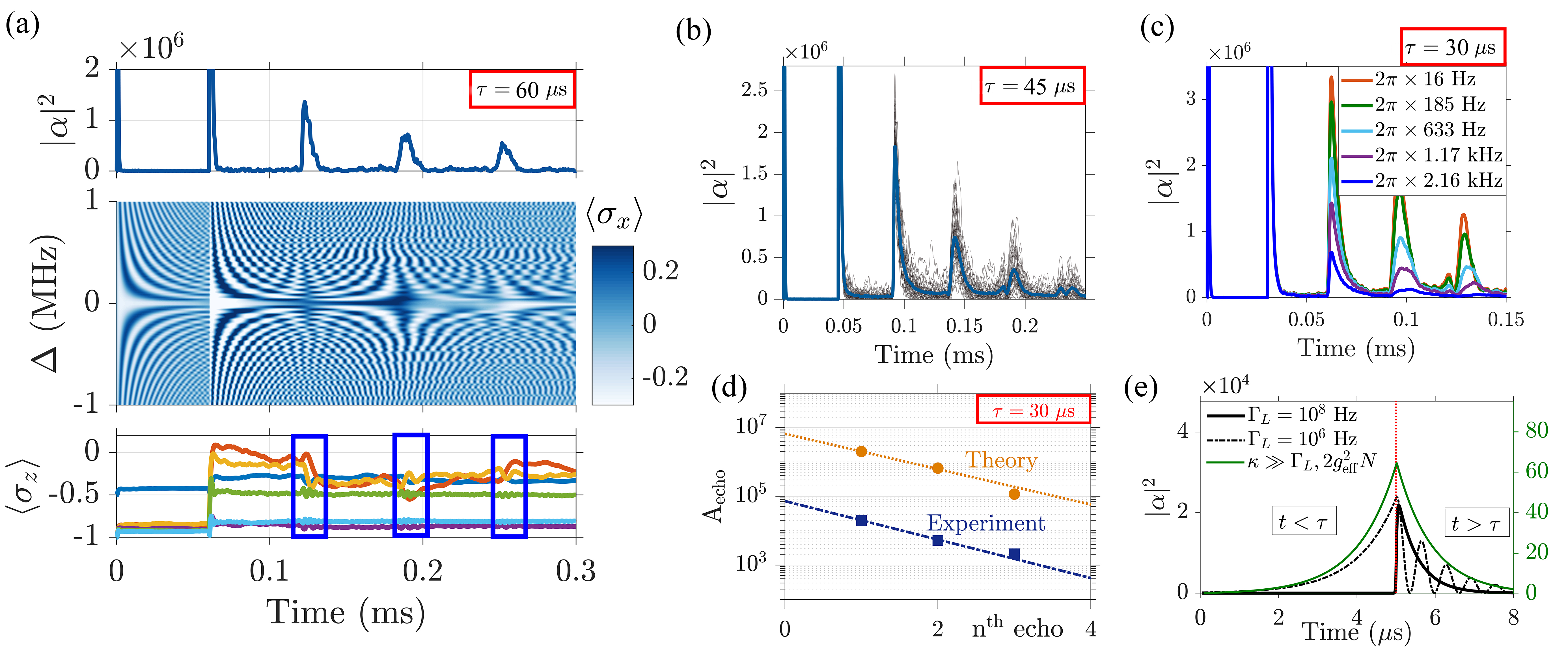}
\caption{Numerical results 
(a) Top: Intra cavity photon number $|\alpha|^2$ as a function of time for a numerical calculation with $\tau=60$ $\mu$s and $\Gamma= 2\pi\times0.5$ kHz (for other parameters, see text).  
Center: $\langle\sigma_x\rangle$ for the spins in a small detuning region as a function of time. 
Bottom: $\langle\sigma_z\rangle$ for selected frequency classes. 
(b) Theoretical results averaged over 45 different samples of the detuning distributions. 
We consider $\tau= 45$ $\mu$s, $\Gamma= 2\pi\times0.5$ kHz and we observe that the echoes change shape over time.
(c) Averaged $|\alpha|^2$ for $\tau= 30$ $\mu$s and different values of the spin dephasing rate $\Gamma$.
(d) Emitted number of photons in the echoes  as a function of their order of arrival according to theory (dotted) and experiment (dot-dashed, arbitrary units). The results are shown for $\tau= 30$ $\mu$s and $\Gamma= 2\pi\times2.5$ kHz.  An exponential fit $y= a e^{-b x}$  reveals a decay rate of $b= 2\pi\times \{6.29$ kHz (theory), $6.84$ kHz (experiment)$\}> \Gamma$. 
(e) The first spin echo signal for different values of the Lorentzian inhomogeneous width $\Gamma_L$ as computed from Eq.~\ref{E7}, assuming an (overestimated) Holstein-Primakoff spin excitation amplitude $\beta = 1$.}
\label{F2}
\end{figure*}

In Fig.~\ref{F2}(a) we present the dynamics of the spin ensemble for $\tau= 60$ $\mu$s, $\Gamma= 2\pi\times 0.5$ kHz. The top panel shows the square of the cavity field amplitude, {\it i.e.}, the intracavity photon number, $|\alpha|^2\equiv  |\langle a\rangle|^2$.
The first two peaks, which extend beyond the border of the figure, correspond to the two external driving pulses, and we observe that the Hahn echo sequence generates multiple echoes separated by $\tau= 60$~$\mu$s. To understand this better, we plot $\langle\sigma_x\rangle$ for single spins in different frequency classes close to the resonance in the center panel of Fig.~\ref{F2}(a), where $\Delta= \delta^j_a-\delta_c$. For simplicity, we consider $\omega_c= \omega_p$ and hence $\Delta= \delta^j_a$. The refocussing of the spins is characterized by  $\langle\sigma_x\rangle$ and $\langle\sigma_y\rangle$ converging to the same finite value for a range of detunings at $2\tau$ and with reduced strength at later multiples of $\tau$. The lower panel of Fig.~\ref{F2}(a) shows the dynamics of the $z$-component of the spin vector for individual spins in few selected frequency classes. The blue boxes highlight the region where the individual spins get a boost due to the strong mean field, appearing in the cavity when the collective spin refocusses. 

In Fig.~\ref{F2}(b) we analyze the shape of the echo signal by averaging multiple trajectories (shown in grey) with randomly sampled Gaussian distributed frequency classes for $\tau= 45$ $\mu$s and $\Gamma= 2\pi\times 0.5$ kHz. A consistent plot of $|\alpha|^2$ demands many closely spaced frequency classes ($N_k\gg 10^5$), which is a challenging task even with a mean field approximation. Averaging multiple realizations results in a more consistent data and $|\alpha|^2$ averaged over $45$ realizations is shown with the bold curve.
 The rate $\Gamma$, which takes into account dephasing due to different mechanisms such as mutual interactions of the spins, spectral diffusion due to crystal deformation and phonon induced energy shifts, leads to vanishing of $\langle\sigma_x\rangle$ and $\langle\sigma_y\rangle$ and breakdown of refocussing at large $t$. The separation between the echoes increases with increasing $\tau$, as evident from Fig.~\ref{F2}(a-c) and the phase of the output cavity field due to the echoes at $t= 2\tau, 3\tau ..$ depends on the phase of the refocussing pulse~\cite{Supp}.

Fig.~\ref{F2}(c) shows how an increase of the spin dephasing rate $\Gamma$ causes a flattening of the shape of the echoes for $\tau= 30$ $\mu$s. It also explains the symmetric shape of the echo observed in the experiments with the dephasing rate $\Gamma= 2\pi\times 2.5$ kHz (Fig.~\ref{F1}(a)) and in Ref.~\cite{Supp}.

Following Ref.~\cite{arxivpreprint1}, in Fig.~\ref{F2}(d), we plot the emitted photon number in each echo  $A_{\rm echo} = \int_{{\rm echo}} |\alpha|^2 \kappa$ $dt$ as a function of its order of emission from both experiments and numerical calculations with $\tau= 30$ $\mu$s and $\Gamma= 2\pi\times 2.5$ kHz.
The experimental intensity data is not absolutely calibrated and has been displaced (in y-direction) by an arbitrary amount in the plot for better comparison with theory. 
An exponential fit, which is independent of the arbitrary displacement, reveals that the echoes decay faster than the spin dephasing rate $\Gamma$ (see also Fig. S1 (a)-(d) in \cite{Supp}) while we observe an excellent agreement between theory ($2\pi\times6.29$ kHz) and experiment ($2\pi\times6.84$ kHz).
We attribute this faster decay to the incomplete refocusing of the spins by the weakening pulses associated with the observed distortion and lengthening of the pulse shapes and the reduction in their amplitude. 
The inhomogeneous spin-cavity coupling may also affect the effective non-linear dynamics.

We can understand the shape of the first Hahn echo observed in Fig.~\ref{F2}(b) by a simple analytical model. Rather than solving the complete spin dynamics analytically, we assume that a perfect $\pi/2$ and $\pi$ pulse have been applied at $t=-\tau$ and $t=0$ respectively to all the spins. This implies that right after $t=0$, the spin excited states have acquired a phase of ${\rm exp}(i\Delta_j \tau)$ with respect to the spin ground states. Since the spin excited states evolve as ${\rm exp}(-i\Delta_j t)$, they come in phase at $t=\tau$. To model the resulting Hahn echo pulse shape, we employ the Holstein-Primakoff approximation~\cite{PhysRev.58.1098} and treat all spins as harmonic oscillators prepared in a coherent state of complex amplitude $\beta$ ${\rm exp}(i\Delta_j\tau)$ at $t=0$. 
Assuming $\Gamma= 0$, the mean field equations for the intracavity field operator and the spin lowering operator take the following form in the frequency domain
\begin{equation}
\begin{array}{lll}
-i\omega\langle \tilde{a}(\omega) \rangle &=& -\frac{\kappa}{2}\langle \tilde{a}(\omega) \rangle - i\sum^N_{j}g_j\langle\tilde{\sigma}_j(\omega)\rangle, \\
-i\omega\langle \tilde{\sigma}_j(\omega) \rangle &=& -(\gamma + i\Delta_j)\langle \tilde{\sigma}_j(\omega) \rangle  - ig_j\langle\tilde{a}(\omega)\rangle  \\
& &+ \frac{\beta}{\sqrt{2\pi}} e^{i\Delta_j\tau}.
\label{E5}
\end{array}
\end{equation}
The above equations can be formally solved, which yields 
\begin{align}
\langle \tilde{a}(\omega)\rangle = \frac{\frac{-i}{\sqrt{2\pi}}\sum^N_j\beta g_j e^{i\Delta_j\tau}/(\gamma + i\Delta_j - i\omega)}{\frac{\kappa}{2} - i\omega + \sum^N_j g^2_j /(\gamma + i\Delta_j - i\omega)}.
\label{E6}
\end{align}
Since the detunings $\Delta_j$ have a continuous distribution, the summation sign can be replaced by an integral in the limit of large $N$, i.e. $\sum^N_{j}\cdot\rightarrow N\int f(\Delta) d\Delta$. $f(\Delta)= \frac{\Gamma_L/2\pi}{\Delta^2 + \Gamma^2_L/4}$ for a Lorentzian distribution and $f(\Delta)=\frac{1}{\sqrt{2\pi}\Gamma_G}{\rm exp}(-\Delta^2/2\Gamma^2_G)$ for a Gaussian distribution, where $\Gamma_L$ and $\Gamma_G\sqrt{8\ln(2)}$ are their full width at half maxima (FWHM) respectively. In principle, Eq.~\ref{E6} can be solved for a Gaussian distribution~\cite{PhysRevA.86.063810}, however, it is not possible to obtain a general analytical expression for $\langle a(t)\rangle$, and we shall therefore provide analytical results for Lorentzian distributions. Assuming that the inhomogeneity in the light-matter coupling is weak, such that $g_j\rightarrow g_{\rm eff}= \sqrt{\frac{1}{N}\sum^N_j|g^2_j|}$, and  putting $\gamma= 0$,  the above equation can be transformed to the time domain, leading to  
\begin{equation}
\begin{array}{lll}
\langle a(\boldsymbol{t<\tau})\rangle &=&\Big(\frac{2i\beta N\Gamma_L}{\sqrt{2\pi}}\Big)\Big[\frac{e^{-\Gamma_L(\tau-t)/2}}{-\Gamma^2_L - \kappa\Gamma_L - 2g^2_{\rm eff}N}\Big] \\ \\
\langle a(\boldsymbol{t>\tau})\rangle &=&\Big(\frac{2i\beta N\Gamma_L}{\sqrt{2\pi}}\Big)\Big[\frac{4e^{\Sigma_-(\tau-t)}}{\Theta_+} - \frac{4e^{\Sigma_+(\tau-t)}}{\Theta_-}\Big],
\label{E7}
\end{array}
\end{equation}
where $\zeta^2= 16g^2_{\rm eff}N - (\Gamma_L - \kappa)^2$, $\Sigma_\pm= (\Gamma_L + \kappa \pm i\zeta)/4$, and $\Theta_\pm= \zeta(3i\Gamma_L + i\kappa \pm \zeta)$. 
We use Eq.~\ref{E7} to plot $|\alpha|^2$ in Fig.~\ref{F2}(e) for different values of $\Gamma_L$. 
It is evident from Eq.~\ref{E7} that for $t<\tau$, the refocussing of the spins lead to $\langle a \rangle\propto e^{-\Gamma_L(\tau-t)/2}$ for any choice of parameters. 
However, for $t>\tau$, the choice of parameters governs the shape of the decaying echo signal. When $\Gamma_L$ is sufficiently large, i.e. $(\Gamma_L-\kappa)>4g_{\rm eff}\sqrt{N}$, $\Sigma_\pm$ is real and $\langle a \rangle$ is proportional to $(2e^{-\kappa(t-\tau)/2}-e^{-\Gamma_L(t-\tau)/2})$. 
If $\Gamma_L>\kappa$ (bold blue curve), the second term vanishes rapidly and the shape of the pulse is predominantly dictated by the decay of the field out of the cavity i.e. $\langle a \rangle \propto e^{-\kappa(t-\tau)/2}$. 
When the collective coupling $g_{\rm eff}\sqrt{N}$ is sufficiently large, $\zeta$ becomes real and exhibits damped oscillations (dot-dashed orange curve) due to the coherent exchange of energy between the spins and the cavity mode. 
On the other hand, when the spins are coupled to a bad cavity, i.e. $\kappa\gg\Gamma_L, 2g^2_{\rm eff}N$, the echo shape becomes symmetric about $t=\tau$ (green bold curve, right hand axis). 
Note that the above analysis assumes $\gamma=\Gamma=0$, and as we observed above, the echo pulses from a good cavity may also be symmetric if $\Gamma$ is large.

\emph{Conclusion--} To conclude, we have observed and characterized the appearance of echo trains after the simple Hahn echo sequence, and we have argued and shown by theoretical calculations that they are due to spin refocussing by previous echo pulses. 
Our mean field calculations show that the echo field amplitude is indeed strong enough to appreciably alter the individual spin states, and hence cause their refocussing at later times.
Due to the finite duration of the pulse, its refocussing effect on differently detuned spin components is more complex, and the later echo is weaker and has a more complex structure.
Note that the mechanism leads to even later echoes which may all carry contributions from the refocussing by both the most recent and earlier echo pulses.
Our numerical results reproduce most of the qualitative features observed in the experiments and shows excellent quantitative agreement with the observed decay of the integrated echo pulse intensities.

Given the complexity of the model, it is difficult to analytically predict the shape of secondary echoes.  Future efforts shall be devoted to understand the scaling laws that govern the gradual reduction of the self-stimulated echo amplitudes, and how these depend on the spin and cavity parameters. 
It is evident from the present study that in addition to the cavity linewidth and the spin dephasing rates \cite{arxivpreprint1}, this decay will also depend non-linearly on the intensity of the echoes, e.g., due to the incomplete refocussing by previous echoes, and the total number of spins effectively coupled to the resonator. Other factors such as the distribution of detunings and coupling strengths may offer varying contribution to the precise behavior of this decay. Although the secondary echoes studied here may constitute an unavoidable feature to be mitigated in quantum memory or transducer protocols involving spin ensembles, they may also offer opportunities in ESR spectroscopy. Aside from the potential to improve signal-to-noise by averaging multiple echoes in one train, these secondary echoes could provide an efficient route to obtaining information on loss rates in the system~\cite{arxivpreprint1} or reveal spectroscopic information, for example in more complex spin systems with unresolved hyperfine couplings. The understanding and theoretical framework that we present here forms a basis to explore such opportunities in more detail.

\nocite{PhysRevLett.113.154101}.

\emph{Acknowledgements--} K.D and K.M acknowledge support from the European Union FETFLAG program, Grant No. 820391 (SQUARE) and the Villum Foundation. G.D and J.J.L.M. acknowledge support from the Engineering and Physical Sciences Research
Council (EPSRC) through the Centre for Doctoral Training in Delivering Quantum Technologies
(EP/L015242/1) and
QUES2T (EP/N015118/1), as well as from the Horizon 2020 research and innovation programme through grant agreement No. 771493 (LOQO-MOTIONS).

\bibstyle{apsrev4-1}

\newpage

\end{document}


\begin{center}
\begin{large}
\textbf{Supplemental material for \\ Self-stimulated pulse echo trains from inhomogeneously broadened spin ensembles}
\end{large}
\begin{center}
Kamanasish Debnath\hspace{0.8mm}$^1$, Gavin Dold~$^{2,\hspace{0.6mm}3}$, John J. L. Morton~$^{2,\hspace{0.6mm}4}$ and Klaus M{\o}lmer~$^1$
\end{center}


$^1$ \textit{Department of Physics and Astronomy, Aarhus University, Ny Munkegade 120, DK-8000, Aarhus C, Denmark}

$^2$ \textit{London Centre for Nanotechnology, University College London, London WC1H 0AH, United Kingdom}

$^3$ \textit{National Physical Laboratory, Hampton Road, Teddington TW11 0LW, United Kingdom}

$^4$ \textit{Department of Electronic and Electrical Engineering, UCL, London WC1E 7JE, United Kingdom}
\date{\today}
\end{center}

\section{Experimental Setup}

The experiment presented in Fig.~1 of the main text used a sample of 200 ppm $^{145}$Nd:Y$_2$SiO$_5$, driven by a planar NbN superconducting resonator fabricated on its surface. It was cooled to 13 mK in a dilution refrigerator (BlueFors LD400) with a magnetic field supplied via a vector magnet and measured in transmission via two weakly coupled antennae. Continuous-wave (CW) measurements of the resonator indicate a zero-field resonance frequency $\omega_\mathrm{c} = 2\pi \times 8.071 \textrm{ GHz}$ and quality factor $Q = 72,000$. A magnetic field applied close to the D$_1$ optical axis brings the $m_\mathrm{I} = +\tfrac{7}{2}$ ESR transition of $^{145}$Nd in resonance at $B_0 = 326 \textrm{ mT}$, where an avoided crossing is observed. We estimate that 50\% of the coupling strength comes from $1.2 \times 10^{10}$ spins within 4 $\mu$m of the resonator.  Analysis of the avoided crossing observed in CW yields a cavity linewidth (FWHM) $\kappa = 2\pi \times 153.8 \textrm{ kHz}$, inhomogeneous linewidth $\Gamma_\mathrm{inh} = 2\pi \times 5.98$  $\mathrm{MHz}$ and a collective coupling strength $g_\mathrm{eff} = 2\pi \times  5.933 \textrm{ MHz}$. This leads to a high cooperativity $C = \frac{4g_\mathrm{eff}^2}{\kappa \Gamma} = 153$. Measurement of coherence time via a Hahn echo sequence yields $T_2 = 409 \pm 14 \textrm{ }\mu\textrm{s}$, corresponding to a homogeneous linewidth $\gamma = 2\pi \times 389 \textrm{ Hz}$.



The spin dynamics of the inhomogeneously broadened ensemble were investigated by performing pulsed spectroscopy using a custom-built ESR spectrometer. Pulses at 8 GHz are generated by modulating a vector signal generator (VSG) using an arbitrary waveform generator. Coaxial cables carry the pulses to the 13 mK stage of the fridge, attenuated by 50 dB to reduce thermal noise at the sample. The returned signal is amplified with a cryogenic low-noise amplifier, through a fast RF switch which protects the spectrometer from high power pulses, and demodulated with an I/Q mixer referenced against the VSG. 

\begin{figure}
\centering
\includegraphics[width=\textwidth]{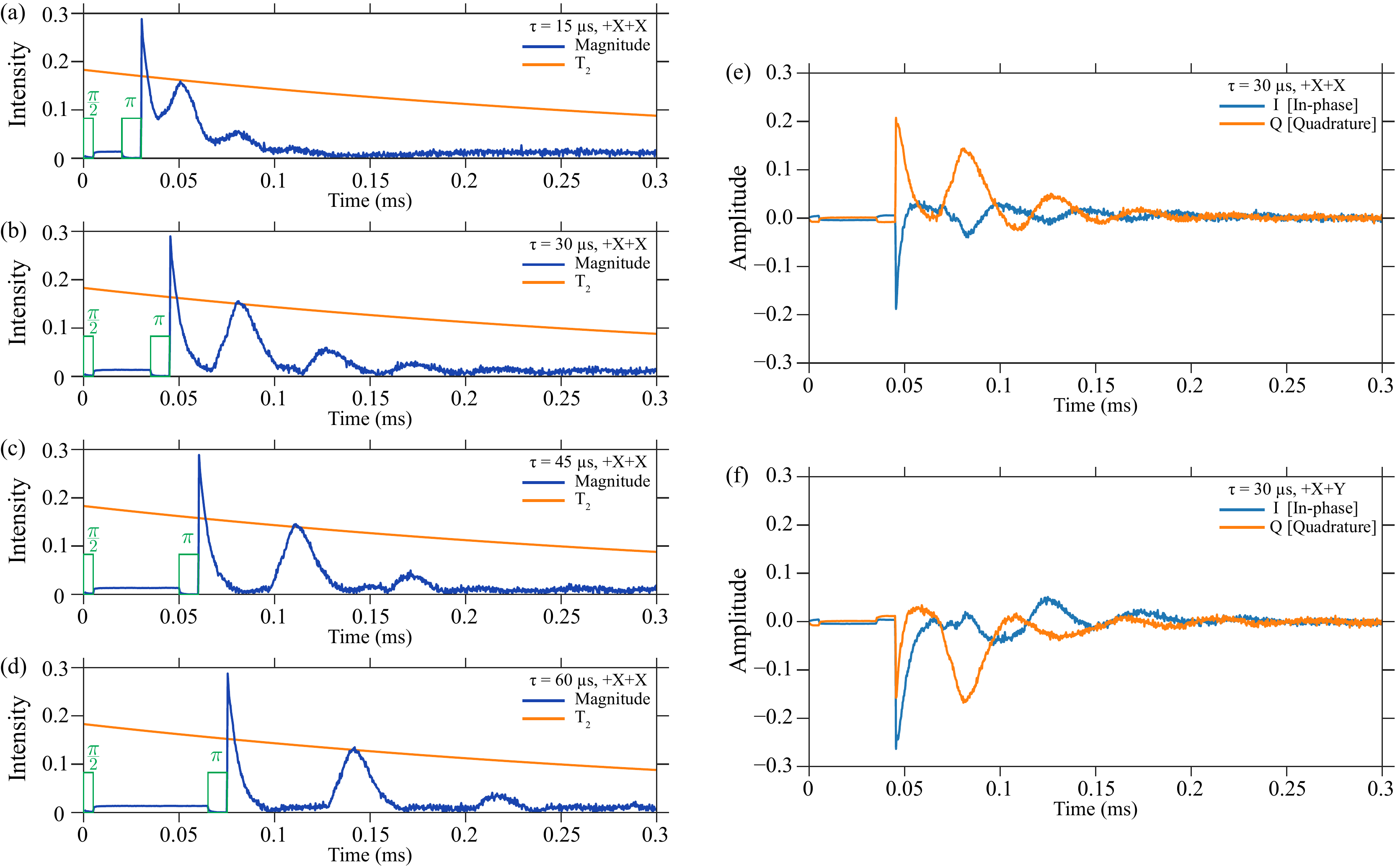}
\caption{Experimental data of a Hahn echo sequence on the $^{145}$Nd spin ensemble, producing an echo train. (a--d) Varying the wait time $\tau$ = \{15, 30, 45, 60\} $\mu$s increases the spacing between subsequent echoes in the train. Subsequent echoes decay significantly faster than the $T_2$ determined from integrating the primary echo. (e, f) Changing the phase of the refocusing pulse \mbox{\{+X, +Y\}} with respect to the initial pulse [+X] changes the phase of the echoes within the train.}
\label{Supp_ExptData}
\end{figure}

\section{Experimental Results}
Fig.~\ref{Supp_ExptData} below shows the intensity and amplitude of the output field following a conventional Hahn echo sequence of two square pulses with $\pi$-pulse length of 10 $\mu$s. We vary $\tau$ over \{15, 30, 45, 60\} $\mu$s and plot the magnitude $\sqrt{\mathrm{I}^2 + \mathrm{Q}^2}$ of each acquired sample in Fig.~\ref{Supp_ExptData}(a-d). Following the pulses, the fast RF switch opens at 45 $\mu$s, resulting in a spike in voltage as the resonator continues to ring down. The time between subsequent echoes in the echo train is seen to increase in accordance with the increased $\tau$. The data presented in Fig.~1 of the main text are taken using $\tau = 30$ $\mu$s.

\begin{figure}
\centering
\includegraphics[width=0.7\textwidth]{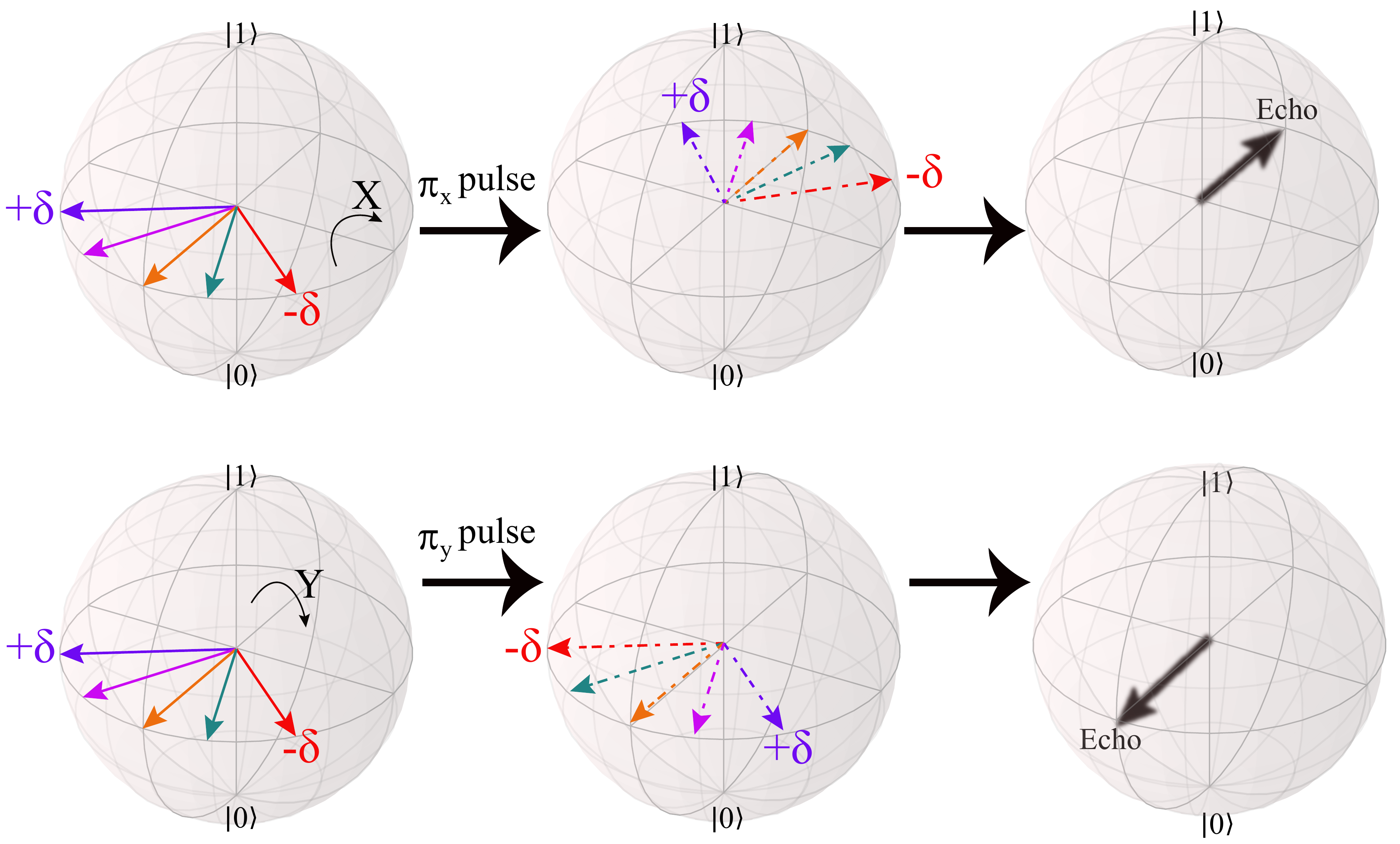}
\caption{Schematic showing the shift of the refocussing point depending on the phase of the $\pi$ pulse.}
\label{Supp_referee}
\end{figure}

Changing the phase of the refocusing pulse \{+X, +Y\} with respect to the phase of the initial pulse [+X], for a fixed $\tau = 30$ $\mu$s, produces the plots in Fig.~\ref{Supp_ExptData}(e, f). Here the channels I and Q are plotted individually so the phase of the echo is visible. A [+X+X] sequence produces a train of echoes with similar phases, while the [+X+Y] sequence in Fig.~\ref{Supp_ExptData}(f) inverts the phase of the initial echo and may further affect the phase of subsequent echoes in the train. 
This is because changing the phase of the refocussing pulse leads to rotation of the spins around a different axis, thereby causing the refocussing to occur along a different direction on the Bloch sphere. This may lead to a change of sign or a mixing of the quadratures of the output cavity field. A schematic of this mechanism is shown in  Fig.~\ref{Supp_referee} for the perfect Hahn echo sequences with different phases of the $\pi$ pulse.

\section{Effect of $\tau$, $N_k$ and Superradiance}
\begin{figure}[h!]
\centering
\includegraphics[width=0.95\textwidth]{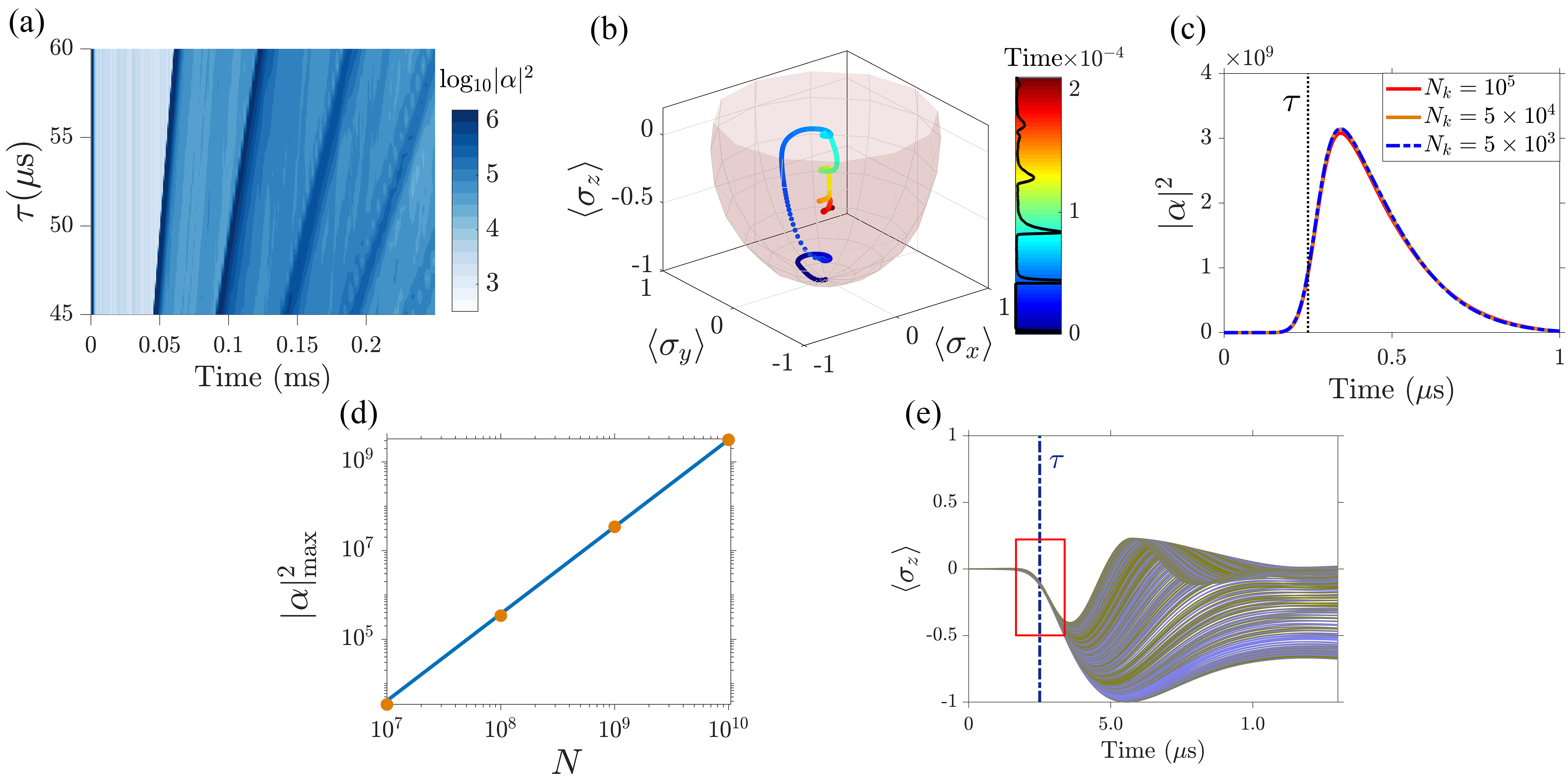}
\caption{Numerically calculated (a) $|\alpha|^2$ (averaged over 45 realizations) as a function of $\tau$ for $\Gamma= 2\pi\times 2.5$ kHz.
(b) Dynamics of a single spin with detuning $5\times 10^4$ Hz and $\Gamma= 2\pi\times 2.5$ kHz on a Bloch sphere representation. The colorbar shows the time and the corresponding curve of $|\alpha|^2$ is superimposed on the colorbar with identical time axis.
(c) The superradiant pulse due to collective decay of the spins using the analytical model described in the main text for different values of $N_k$ and $\beta=1$.
(d) The maximum intracavity photon number $|\alpha|_{\rm max}^2$ scales quadratically with the number of spins $N$.
(e) $\langle \sigma_z \rangle$ of spins in different frequency classes showing evidence of collective decay around $t = \tau$ using the analytical model derived in the main text.
Parameters considered in panel (c)-(e) are $\kappa= 2\pi\times 150$ kHz, $g= 2\pi\times 8$ Hz, $\gamma=\Gamma=0$, $N= 10^{10}$ (panel (c) and (d)) and  $\beta= 1$.}
\label{Supp_F1}
\end{figure}

As shown in the main text, the secondary echoes at $3\tau, 4\tau..$ and so on are stimulated by the first echo at $2\tau$ with possible contributions due to refocussing by later pulses.
In Fig.~\ref{Supp_F1}(a), we show the dynamics (averaged over $45$ realizations) of the spins after the conventional Hahn echo pulse sequence as a function of the time delay $\tau$. 
With increasing time interval $\tau$, the separation between the echoes increases as shown in Fig.~\ref{Supp_F1}(a). 
This provides sufficient evidence that the observed peaks in $|\alpha|^2$ is a result of the spins undergoing \textit{de}-focussing and \textit{re}-focussing after every time interval $\tau$.

Fig.~\ref{Supp_F1}(b) shows the Bloch vector evolution for a single spin with detuning $\delta^i_a = 5\times 10^4$ Hz. 
The value of $|\alpha|^2$ is superimposed on the colorbar with the color code representing the time axis.  
The first pulse excites the spin from the ground state to a superposition state, which is followed by its precession with frequency $\delta^i_a$ in the $\langle\sigma_x\rangle$-$\langle\sigma_y\rangle$ plane. 
The second pulse excites the spin further, where the spin precesses for $\tau= 45$~$\mu$s before the first echo is emitted.
Clearly, imperfect $\pi/2$ and $\pi$ pulses can generate a Hahn echo. 
The echo pulse also induces rotation of the Bloch vectors, which is evident from the sudden change of the z-components of the Bloch vector in the top part of Fig.~\ref{Supp_F1}(b). 
This process repeats and is the cause of several later spin echoes until either dissipative dephasing or the growing duration of the pulses due to the complex Rabi dynamics make their amplitude too weak to appreciably rotate the Bloch vector and cause further echoes.

As all spins come back in phase when they are subjected to refocussing around $t=\tau$, the collective dipole interacting with the cavity field scales linearly with the number of spins, and in analogy with classical antenna arrays we may expect an emitted power proportional with $N^2$. Since the cavity plays an important role and couples to the spin dynamics in a non trivial manner, cf., the possibility to observe synchronization of inhomogeneously broadened spin ensembles in cavities~\cite{sync_KD, Holland_prl}, we investigate the emission in more detail.     
Assuming that a perfect $\pi/2$ and $\pi$ pulse has been applied at $t=-\tau$ and $t=0$ respectively, the spins are initialized in a superposition state with a detuning dependent excitation amplitude  $\beta {\rm exp}(i\Delta_j\tau)$, such that all the spins refocus at $t= \tau$, leading to an echo and collective emission.
Fig.~\ref{Supp_F1}(c) shows the time dependent $|\alpha|^2$ for different values of $N_k$, showing that the emission of radiation is almost independent of the discretization detunings as long as $N_k$ is reasonably large.
In Fig.~\ref{Supp_F1}(d), we plot the maximum emission $|\alpha|^2_{\rm max}$ as a function of the number of spins $N$. A fit of the data, $y= a N^b$, with $b= 1.96$, confirms the superradiant emission characteristics, expected also from classical, non-interacting dipoles.
In Fig.~\ref{Supp_F1}(e), we plot $\langle\sigma_z\rangle$ for spins in selected frequency classes as a function of time around the echo time $t=\tau$.
In the time interval around the refocussing time $\tau$, indicated by the red box, the spin excitation decays in agreement with the power emitted into the superradiant pulse. The collective behaviour mediated by the cavity and the behaviour of the spins becomes more complex in the subsequent echo pulses due to the pulse duration, imperfect refocussing, inhomogeneities, etc.